*Article*

# MEDNC: Multi-ensemble deep neural network for COVID-19 diagnosis

Lin Yang [1], Shuihua Wang [1] and Yudong Zhang[1,*],

[1] School of Computing and Mathematical Sciences, University of Leicester; ylin131@gmail.com, shuihuawang@ieee.org, yudongzhang@ieee.org

* Correspondence: Yudong Zhang

**Abstract:** Coronavirus disease 2019 (COVID-19) has spread all over the world for three years, but medical facilities in many areas still aren't adequate. There is a need for rapid COVID-19 diagnosis to identify high-risk patients and maximize the use of limited medical resources. Motivated by this fact, we proposed the deep learning framework MEDNC for automatic prediction and diagnosis of COVID-19 using computed tomography (CT) images. Our model was trained using two publicly available sets of COVID-19 data. And it was built with the inspiration of transfer learning. Results indicated that the MEDNC greatly enhanced the detection of COVID-19 infections, reaching an accuracy of 98.79% and 99.82% respectively. We tested MEDNC on a brain tumor and a blood cell dataset to show that our model applies to a wide range of problems. The outcomes demonstrated that our proposed models attained an accuracy of 99.39% and 99.28%, respectively. This COVID-19 recognition tool could help optimize healthcare resources and reduce clinicians' workload when screening for the virus.

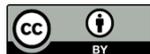

## 1. Introduction

The ongoing COVID-19 epidemic is one of the most devastating health crises in recent history. COVID-19 is still considered a public health emergency of international concern (PHEIC) by the World Health Organization Director-General in January 2023[1]. Despite global improvements since the height of the Omicron transmission a year ago, over 170,000 deaths have been attributed to COVID-19 in only the previous eight weeks[1]. The reproduction number for COVID-19 during the early months of the pandemic ranged from 2.24 to 3.58[2], which means that each infected individual, on average, transferred the sickness to two or more persons. This will cause the number of instances to increase from a few hundred in January 2020 to over 672 million in January 2023[3]. It takes an average of 5.1 days for the vast variety of COVID-19 symptoms to appear following infection. Frequent symptoms include fever, dry cough, and fatigue. Headache, hemoptysis, diarrhea, dyspnea, and lymphopenia are further symptoms [4].

Two primary methods of screening for COVID-19 are reverse transcription-polymerase chain reaction (RT-PCR), and chest computed tomography (CT) scans[5]. RT–PCR is a way to find out if a pathogen, like a virus, has specific genetic material [6]. While in CT imaging, a sensitive detector is used in conjunction with collimated X-ray beams, gamma rays, ultrasonography, and other radiation to scan a body part in portions. Then, slides of your bones, blood vessels, and soft tissues are created on a computer[7]. In contrast to the nose test used in RT-PCR[8], which can detect COVID-19 infections, chest CT scans image the patient's lungs using tomography[9]. There have been cases where PCR testing came back negative but patients were verified Covid positive with CT scans[10], suggesting that Chest CT has a higher sensitivity for the detection of COVID-19 compared to RT-PCR. With a sensitivity of over 97% and a specificity of roughly 25%, chest CT has been shown in research to be an accurate diagnostic tool for COVID-19[11]. Chest CT has been increasingly employed as an alternative to RT-PCR in the clinical pathway for identifying COVID-19.

However, it might take plenty of time and work for clinicians to evaluate disorders using CT scans in the traditional approach. Thankfully, deep learning for medical image processing may speed up diagnoses and decrease doctor effort. Deep learning is an artificial intelligence frontier that is inspired by the human brain[12]. Deep learning may be able to replace physicians' years of experience and careful examination in detecting a patient's illness[13]. Patients may quickly gain a perspective that is less biased after an assessment[14]. Professionals have been able to quickly detect patients with potential disease infection thanks to the use of deep learning technologies[15].

Recognition of COVID-19 has been achieved using a variety of distinct deep-learning models. Researchers in Ref [16] proposed a ResNet50 model to detect COVID-19 from chest CTs. Rather than chopping off parts of the image, they feed the entire image's wavelet coefficients into the ResNet's foundational model. The outcome was 92.2% accurate. To detect COVID-19 patients, in Ref [17], five deep CNN learning models were compared. The researchers improved the classification efficiency of all five models by combining conventional picture augmentation with CGAN. The results revealed that ResNet50 had the highest degree of accuracy at 82.9%. In Ref[18], eight previously trained models were compared for their ability to identify COVID-19 patients: According to the findings, DenseNet201 performed best, with an accuracy of 85%. When comparing COVID-19 CT scans to those of other patients with pneumonia or healthy individuals, a CNN design based on SqueezeNet was proposed (Ref [19]). As a result of the design, they were able to achieve an accuracy of 85.03%. Ref [20] developed FCONet using a pre-trained CNN model as its backbone. The data shows that FCONet, on average, was nearly perfect. A CT dataset consisting of 361 pictures was used to train a modified version of a pre-trained AlexNet model, as described in Ref [21]. According to the results, the customized CNN model had the highest accuracy, at 94%.

Our paper's contributions are as follows：
- We propose three deep-learning neural networks named MEDNC (FFC-MEDNC, FCO-MEDNC, FO-MEDNC, FFCO-MEDNC) for recognizing COVID-19 with CT scans
- The accuracy of our proposed FFCO-MEDNC has reached 98.79% and 99.82% on two COVID-19 datasets.
- Our models have also been successful in predicting blood cell and brain tumor images with accuracies of 99.28% and 99.39%, respectively.

This paper's remaining sections are organized as follows: The second section addresses materials and processes. Section 3 contains the findings. Section 4 contrasts the outcomes to contemporary methods. This study concludes in section 5.2.

**2. Materials and Methods**

This segment concentrates on the development and implementation approach for the MEDNC model. We offer approaches for using DL to differentiate between chest CT scans for COVID-19 illnesses and non-COVID-19. Figure 1 is a graph depicting a flowchart that includes these terms.

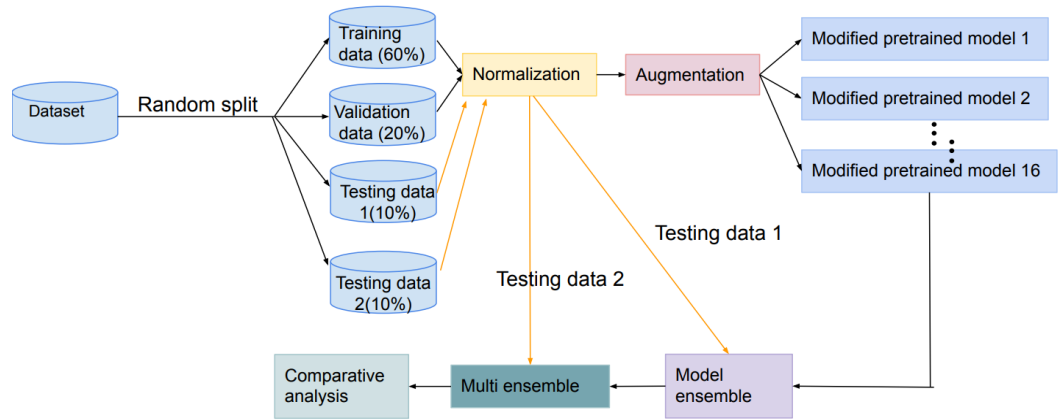

**Figure 1.** Design flowchart for the COVID-19 identification system

*2.1. The Dataset*

The patient's lungs can be seen clearly in a series of images taken using the CT scan technique on the chest. Images in a CT scan series may or may not show diseased areas; for example, the lung is enclosed at the start and finish of each image series. For COVID-19 anomaly detection, it is necessary to have data samples in which lung internal structures can be seen. Only pictures from the center point of the CT sequence are accessible for use, researchers have discussed methods for autonomously selecting images that are visible on CT scans of the lung[22].

To train a deep learning model using data from a DICOM-formatted CT image sample, the user can perform the following Python code to convert DICOM to PNG. Utilize the dicom.read_file() function to initially load the DICOM image. Then, transform the intercept and slope from the DICOM image header. Present the image utilizing the window at 1500 level at -600 and width information from the image header. Lastly, cv2.convertScaleAbs() function is utilized to convert the DICOM data to PNG format. The following two sections include Covid-19 datasets in PNG format[23].

2.1.1. Covid-19 Dataset A

This work employs a CT scan dataset dubbed SARS-CoV-2 [24] for the COVID-19 recognition challenge. From hospitals in Sao Paulo, Brazil, it gathers 2481 CT scan pictures of people of both sexes. COVID-19 positivity was found in 1252 of the CT images and was not detected in 1229 (not normal).

The resolution of these PNG files, which were created from CT scans, ranges from 104 to 416 pixels on the longest dimension. To guarantee a perfectly balanced dataset, we chose 1229 pictures to represent each of the two categories. One CT scan of a patient with COVID-19 from this dataset is shown in Figure 2a; the arrows point to contaminated tissue. Figure 2b shows a contrast CT scan from a patient who did not have COVID-19. The CT image data used in this analysis is listed in Table 1.

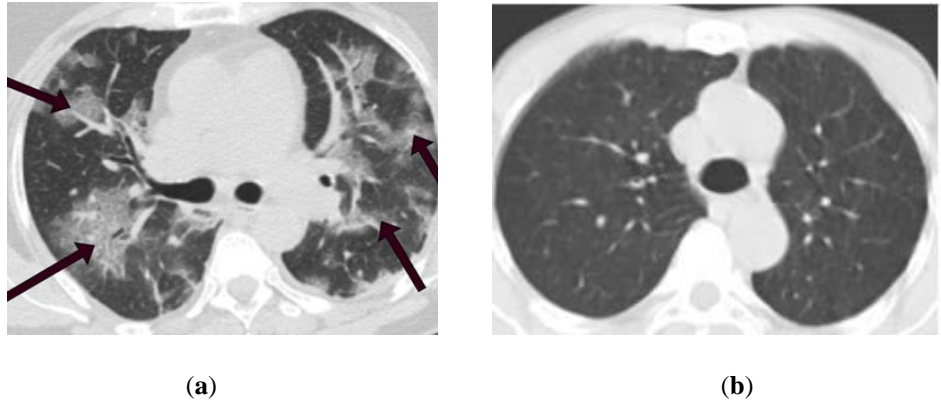

**Figure 2.** Patients with and without the COVID-19 virus in their CT scans. (a) CT scan of a subject with COVID-19. (b) A CT scan of an abnormal (not COVID-19) subject.

**Table 1.** Details about the SARS-CoV-2 dataset of chest CT scan images that were chosen

| Classes | Quantities of Samples | Format |
|---|---|---|
| COVID-19 | 1229 | PNG |
| Non-COVID-19 | 1229 | PNG |

2.1.2. Covid-19 Dataset B

As can be seen in Figure 3, this work applies the suggested MEDNC models to another publicly available CT dataset called COVIDx CT-2A [25]. Random samples of CT images in PNG format with sequential numbers were drawn from both COVID-19/non-COVID-19 datasets. Data from CT images are listed in Table 2.

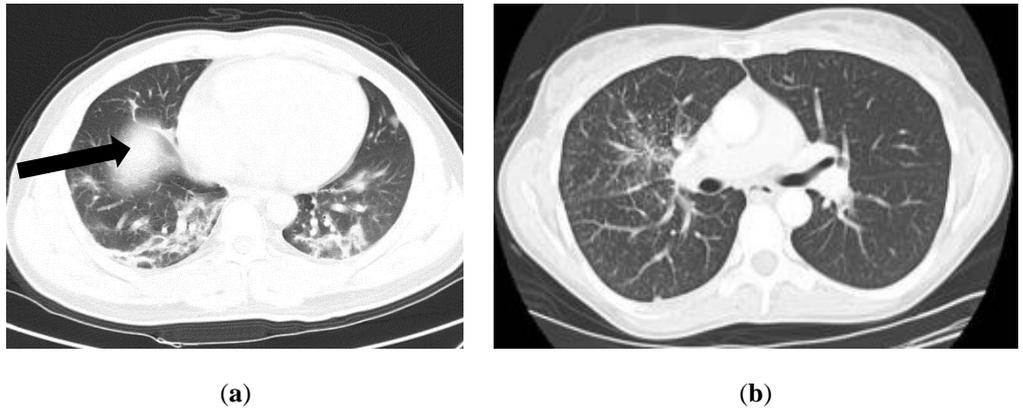

**Figure 3.** Patients with and without the COVID-19 virus in their CT scans. (a) a subject with COVID-19. (b) n abnormal (not COVID-19) subject.

**Table 2.** Data from the COVIDx CT-2A dataset of chest CT scan images that were chosen

| Class | Quantities of Samples | Format |
|---|---|---|
| COVID-19 | 349 | PNG |
| Non-COVID-19 | 349 | PNG |

*2.2. Data Preprocessing*

To begin, a random 60%, 20%, 10%, and 10% split is made between the training, validation, testing A, and testing B sets from the specified dataset. Second, to accommodate the pixel-value representation necessary for image processing, we normalize the range of pixel values from (0, 255) to (0, 1) to train the deep learning model correctly[26]. It can be characterized as follows:

$$x_i' = \frac{x_i - \min(x)}{\max(x) - \min(x)}, \quad (1)$$

where min and max are the initial pixel values of 0 and 255, respectively. x is the entire image, while $i$ is a single pixel within that image. Every dataset used for training, validation, and testing undergoes this rescaling of pixel values.

Moreover, as the input size to the deep learning network must be fixed, all image sizes are rescaled to 224 by 224[27]. It has been found that bigger datasets can improve deep learning classification accuracy. However, it is not always feasible to have a big dataset[28]. Thus, to expand the amount of data without gathering fresh pictures, an augmentation method is utilized[29]. Images from a dataset are enhanced in this work by geometric transformations including rotation and flipping.

*2.3 The Proposed Deep Learning Framework*

In this section, we proposed a multi-ensemble deep learning neural network for COVID-19 (MEDNC) and compared four deep learning models under MEDNC for COVID-19 lung CT recognition.

2.3.1. MEDNC framework

MEDNC is a framework inspired by CNN. It consists of a feature extractor and a classifier, as well as ensemble learning. The feature extractor is constructed from convolutional and pooling layers. A flatten layer, a fully connected layer, and an output layer compose the classifier. The proposed multi-layer ensemble method consists of four potential ensemble models that can improve the accuracy of COVID-19 screening in comparison to the use of individual models.

In this paper, we use pre-trained feature extractors for easy modeling because it may be difficult to train a deep learning model for recognizing COVID-19 infection from scratch due to a lack of CT scans of COVID-19 subjects. The transfer learning method and several different pre-trained models address this problem[30]. One of transfer learning's key benefits is that it allows data to be trained with fewer samples and in less time [31]. And it allows a newly trained model to benefit from the experience of a previously trained model [32]. Table 3 presents the pre-trained part and the novelty portion of our MEDNC.

Table 3. Pre-trained and suggested parts of the MEDNC

| Feature extractor components | Classifier components | Multi-ensemble technique |
|---|---|---|
| ● Convolutional layers<br>● Pooling layers | ● Flatten layers<br>● Fully connected layers<br>● Output layers | ● FFC ensemble<br>● FCO ensemble<br>● FO ensemble<br>● FFCO ensemble |

| Pretrained | Proposed | Proposed |

When it comes to the COVID-19 recognition task, sixteen popular CNN models are selected for the transfer learning algorithm we customized. These include VGG16, ResNet201, ResNet152v2, ResNet50, ResNet101, ResNet101V2, DenseNet201, MobileNet, MobileNetV3 small, MobileNetV2, XceptionNet, ResNet50V2, InceptionResNetV2, NASNet, and DenseNet169. These models were chosen because of their usefulness in computer vision applications. There have been numerous reports of their effectiveness in medical diagnostics[33, 34].

In order to classify images, the selected sixteen models are already reviously trained on the largest dataset ImageNet[35]. Since a large dataset was used to process the selected sixteen models, their learned weights might be applied to medical image recognition. These sixteen models serve as the foundation for further MEDNC model building. To prevent any data loss and make the most of the features extracted for training COVID-19 tasks, the feature extraction layers are frozen after being ImageNet-optimized. Also, to categorize COVID-19 CT images, we pruned the fully connected layers from the original models and created the new classifier network. The details of the suggested classifier are provided below.

A down-sampled feature map is transformed into a one-dimensional array using a flattened layer; a fully connected layer is built by activating connections between neurons in the preceding and following layers with a Rectified Linear Unit (ReLU) [36]. The issue of vanishing gradients can be resolved with the use of ReLU. The following equation is used to arrive at the answer:

$$f(x) = \begin{cases} 0, for\ x < 0 \\ x, for\ x \geq 0 \end{cases}. \quad (2)$$

To avoid overfitting, a dropout is utilized with a ratio of 0.5 to create a more robust model; A Softmax activation is added to the output layer to determine if a CT scan is a COVID-19 infection or not. In contrast to ReLU, Softmax is typically implemented as a classification algorithm in the final layer of a model [37]. The following equation is one possible way to express it.

$$s(x_i) = \frac{e^{x_i}}{\sum_{j=1}^{n} e^{x_j}},$$
$$i = 1,2,\dots k. \quad (3)$$

The sixteen pre-trained models that were utilized are shown with their revised architectures in Figure 4.

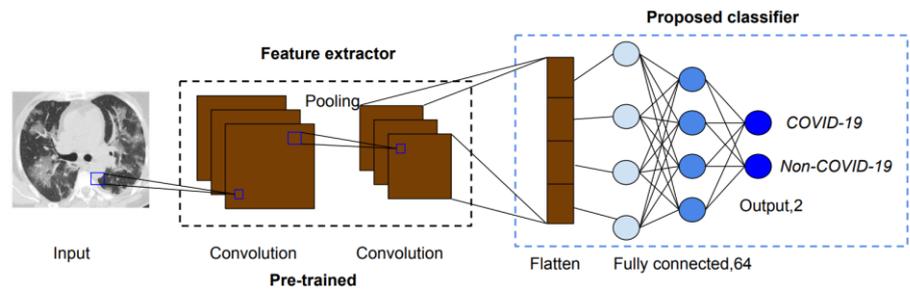

**Figure 4.** The pre-trained feature extractor and the proposed classifier as a whole model

Predicting the course of disease with any degree of precision is essential in healthcare since wrong decisions have serious financial and lifelong consequences. The predictions of separate deep learning classifiers have a high degree of variation, which is a downside when used alone. When asked to classify the same data, the answers from different models might be highly variable because they are all built and trained independently[38]. The multi-layer ensemble of individual models has the potential to solve these issues compared to one layer ensemble suggested before since it reduces variance and is more generalizable than other ensemble techniques and the use of individual models taken alone. Motivated by this fact, we have proposed a MEDNC framework for accurately classifying COVID-19. Table 4 summarizes the MEDNC framework

Table 4. An outline of the MEDNC framework

| MEDNC framework | Levels of ensemble | Configurations |
|---|---|---|
| FFC-MEDNC | Level 2 ensemble | Includes feature extractors ensembles and the fully connected layers ensemble. |
| FO-MEDNC | | Consists of feature extractors ensembles and the output layers ensemble. |
| FCO-MEDNC | | Made up of the fully connected layers ensembles and the output layers ensemble. |
| FFCO-MEDNC | Level 3 ensemble | Includes feature extractors ensembles, the fully connected layers ensembles and the output layer ensemble. |

2.3.2. FFC-MEDNC

One of the MEDNCs we proposed is the feature-fully connected multi-layer ensemble deep neural network for COVID-19 (FFC-MEDNC). It comprises two parts. The initial ensemble combines pre-trained models at the feature level. The second component is an ensemble technique that incorporates feature ensemble model sets at fully connected layers. The next section describes the FFC-DEDNC model's architecture.

Feature Ensemble

The major objective of ensemble models together at the feature level is to group additional input characteristics that a single model is unable to group. Therefore, this feature ensemble layer generates a dataset that combines all necessary CT scan input attributes. Assuming $I_I = \{i_1, i_2, ..., i_n\}$ is the COVID-19 dataset used as input, the following is true:

$$F_x = \{f_{x1}, f_{x2}, ... f_{xn}\}, \qquad (4)$$

where $f_{x1}, f_{x2}, ... f_{xn}$ are the feature derived from the same input $i_n$ by the transfer learning models. In this instance, $n = 2$.

As a result, we can express the ensemble of various pre-trained models as follows:

$$F_e = Concatenate[f_{x1}, f_{x2}, \ldots f_{xn}]. \tag{5}$$

Fully Connected Ensemble

The fully connected ensemble layer brings together the fully connected layers of all proposed CNN models to make fewer trainable parameters of the full ensemble model. The goal of this layer is to make a more accurate probability by putting together the results of all fully connected layers. The output of the pre-trained models' fully connected layer $FC_{xn}$ will be used as a separate input for our model. To finish the FFC-MEDNC model, an output layer was added after the fully connected ensemble layer. For our model, we used the categorical cross-entropy loss function, written as follows:

$$CE = -\sum_i^c t_i \log(f(s)_i), \tag{6}`$$

where $t$ stands for the desired values, $f(s)$ stands for the projected values, and $i$ stands for the label.

In Figure 5 we can see the overall structure of the FFC-MEDNC model.

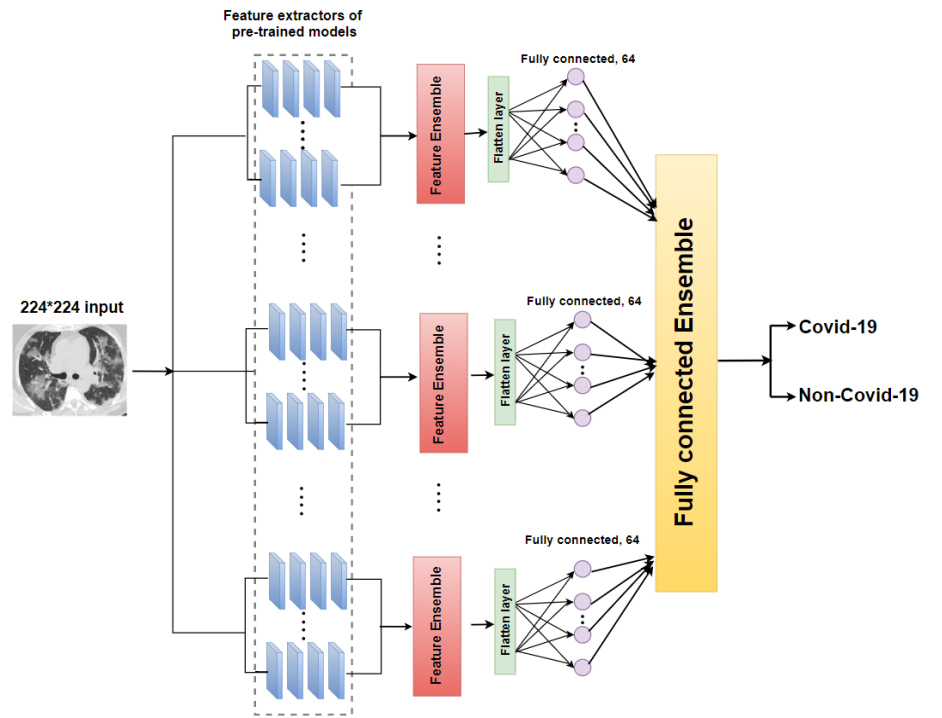

**Figure 5.** The FFC-MEDNC architecture.

2.3.3 FCO-MEDNC

There are two components to the fully connected-output ensemble deep neural network for COVID-19 recognition (FCO-MEDNC). The first component is a fully connected ensemble that employs a combination of pre-trained models. The second section consists of an output ensemble comprising sets of fully connected ensemble models. The architectures of the FFC-MEDNC model are described in the next section.

Fully Connect Ensemble

At the fully connected ensemble, we have groups of neural networks (two in each group) that had pre-trained feature extractors. The principal purpose of the model ensemble operating at the fully-connected level is to aggregate additional input characteristics that the single-model ensemble cannot. In this way, the CT scan input is used by all three sets of models to generate a dataset with all the needed attributes. This FCO-MEDNC model will use the result of the fully connected layer $FC_{xn}$ of the proposed single CNN model as a distinct input. Since we are dealing with a value of $n = 2$, there's this:

$$FC_x = \{fc_{x1}, fc_{x2}, \ldots fc_{xn}\}, \tag{7}$$

To get a more precise likelihood of detecting COVID-19 CT scans, the fully connected layers' outputs are merged into one. Consequently, the ensemble of fully connected layers from various pre-trained models can be expressed as follows:

$$FC_e = Concatenate[fc_{x1}, fc_{x2}, \ldots fc_{xn}]. \tag{8}$$

Output Ensemble

Three fully connected ensemble models are put together at the output layer. This method can be shown in the following way:

$$O_x = \{O_{x1}, O_{x2}, \ldots O_{xn}\}, \tag{9}$$

where $O_{xn}$ is the result of each fully connected ensemble model.

As a result, we can express the ensemble of layers at the output of three fully connected ensemble models as follows:

$$O_e = Concatenate[\{O_{x1}, O_{x2}, \ldots O_{xn}\}], \tag{10}$$

The ensemble model is expected to pick up more features from the combined output to refine its predictions using this approach. Figure. 6 depicts the FCO-MEDNC model's architecture.

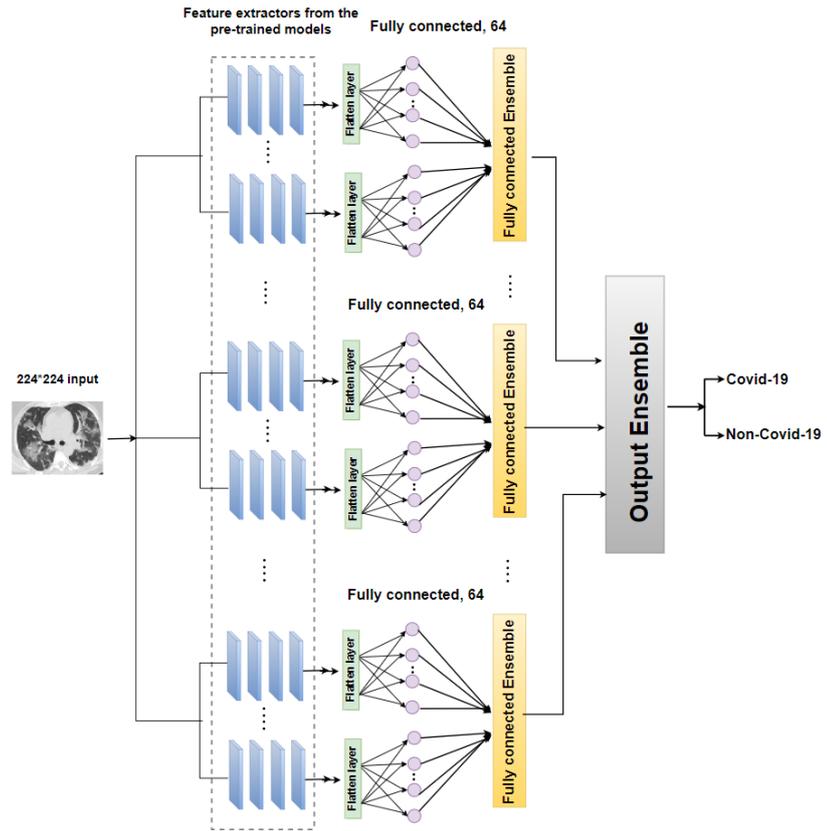

**Figure 6.** The FCO-MEDNC's structural design.

2.3.4 FO-MEDNC

There are two sets of ensemble networks that make up the feature-output ensemble deep neural network for COVID-19 recognition (FO-MEDNC). The first part of the network is similar to the FFC-MEDNC that has an ensemble at the feature level, it mixes feature extractors that have already been pre-trained. The second component is an output ensemble technique, which integrates three different feature ensemble models into a single model. Figure 7 presents a diagrammatic representation of the FO-MEDNC model's internal structure.

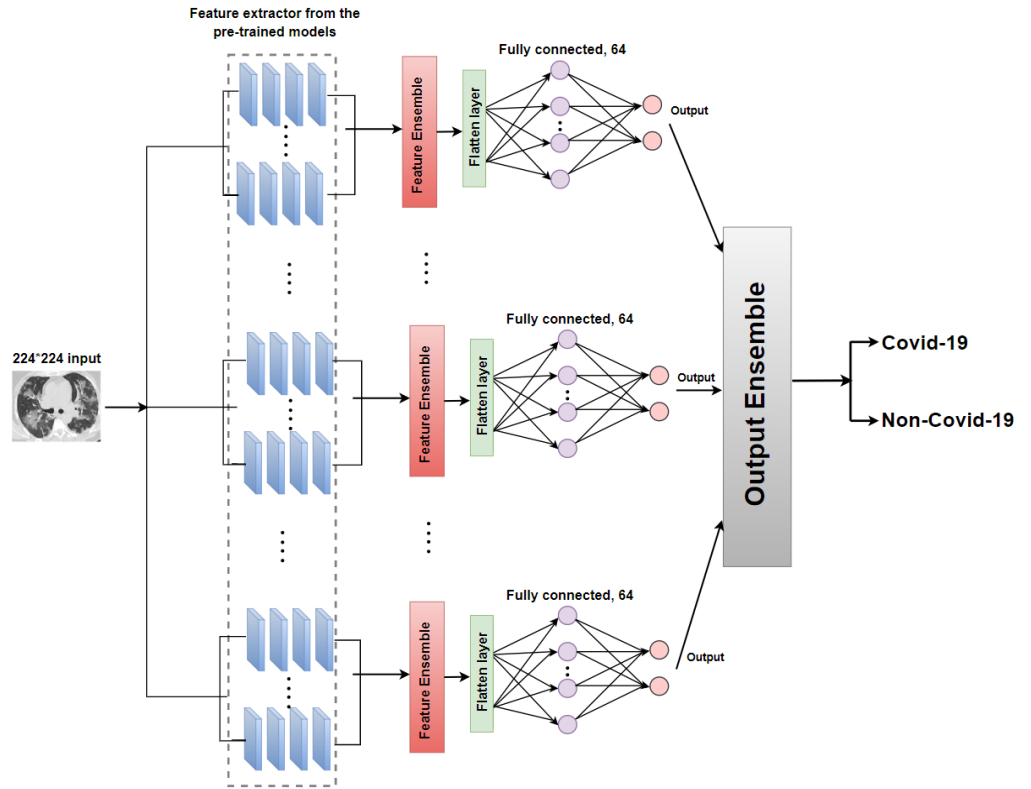

**Figure 7.** The architecture of the FO-MEDNC.

2.3.5 FFCO-MEDNC

Except for the three ensemble methods we described above, we also suggested an extension level 3 for MEDNC. The architecture of the FFCO-MEDNC is shown in Figure. 8. From the figure, we can observe that FFCO-MEDNC has three levels of ensemble sets. First, the feature extractor ensemble is applied. Secondly, the results of the feature extractor ensemble are combined at the fully connected layer. Last but not least, the final level of the ensemble, which is an output ensemble, is used to combine all the input characteristics for COVID-19 screening.

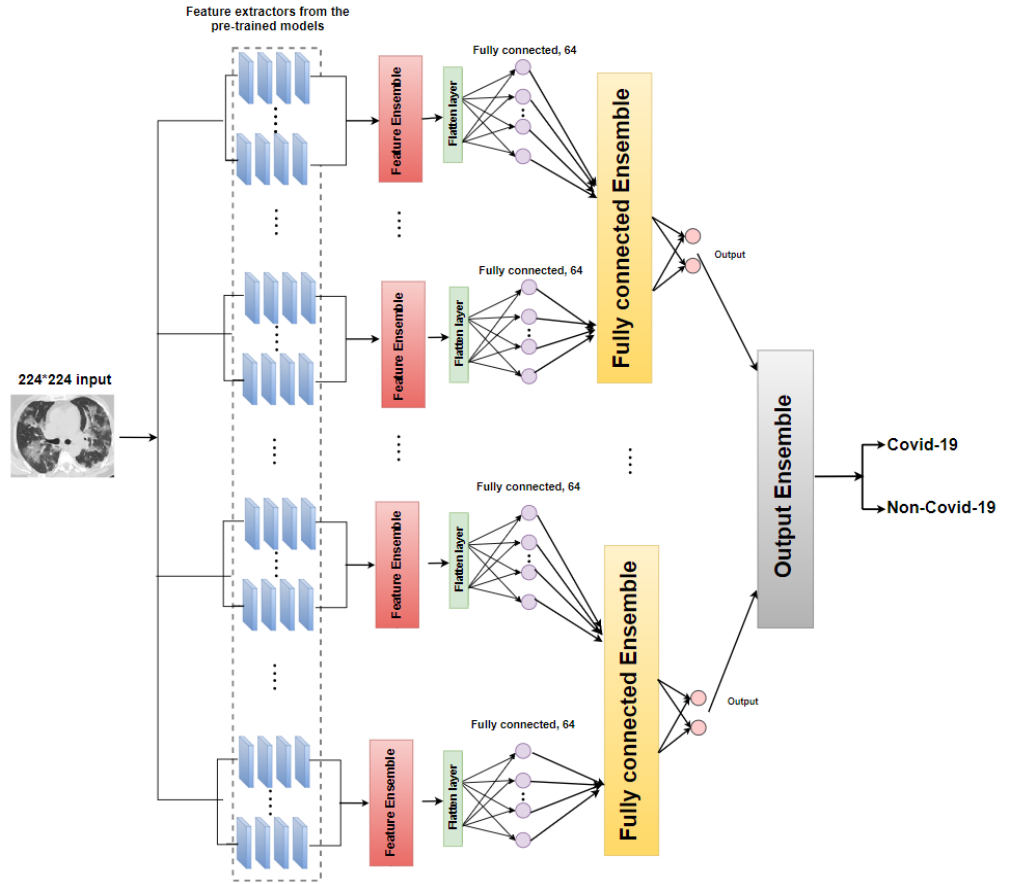

Figure. 8 The architecture of the FFCO-MEDNC

*2.4. Model Evaluation*

2.4.1. Experimental Setup

Python 3.7 (Python Software Foundation, Leicester, U.K.), TensorFlow 2.4.0 (Google, Leicester, U.K.), Keras 2.3.1 (François Chollet, Leicester, U.K.), and Scikit-Learn 0.20.4 (David Cournapeau, Leicester, U.K) are used to create the networks in Jupiter notebook. The models are trained on a desktop PC equipped with an Nvidia GPU RTX2070S (Nvidia, Leicester, U.K.) and an Intel Xeon E5-2680 v2 processor (Intel, Leicester, U.K.) .

2.4.2. Confusion Matrix

Figure 9 depicts a table called confusion matrix, it summarizes the results of a prediction for a classification task[39]. Correct and incorrect estimates are added together and placed in one of four categories. In a True Positive (TP) scenario, both the anticipated and actual outcomes are positive. In a false positive (FP) scenario, the forecast is optimistic whereas the real result is pessimistic. In the case of a True Negative (TN), both the predicted and actual outcomes are negative. A false negative (FN) occurs when the anticipated result is negative but the actual result is positive.

**Actual Values**

|  | Positive (1) | Negative (0) |
|---|---|---|
| **Predicted Values** Positive (1) | TP | FP |
| Negative (0) | FN | TN |

**Figure 9.** A visual depiction of the confusion matrix.

2.4.3. Classification Metrics

The model's performance was measured with the following five indicators[40]:

Accuracy is the proportion of accurate to faulty predictions.

$$\text{Accuracy} = \frac{TP+TN}{TP+FP+TN+FN}. \qquad (11)$$

Precision measures how well a model can identify positive data in a sample.

$$\text{Precision} = \frac{TP}{TP + FP}. \qquad (12)$$

The sensitivity of a model is measured by how well it can identify positive data.

$$\text{Sensitivity} = \frac{TP}{TP + FN}. \qquad (13)$$

F1-Scores is a balanced accuracy and recall measurement.

$$F1 = \frac{\text{Precision} \times \text{Recall} \times 2}{\text{Precision} + \text{Recall}}. \qquad (14)$$

The probability of negative samples identified is counted as specificity.

$$\text{Specificity} = \frac{TN}{TN+FP}. \qquad (15)$$

2.4.4. Monte Carlo Cross-Validation

We use Monte Carlo Cross-Validation as it is more precise than a single validation (hold-out) in evaluating a model's efficacy. Equivalent to repeated random subsampling CV, Monte Carlo Cross-Validation (MCCV) is a method for assessing the quality of a model [41]. Picard and Cook were the first to propose the MCCV approach, which essentially just repeats the holdout technique[42]. The model is trained and validated independently multiple times by randomly splitting the data set into training and validation sets; the results of these validations are then averaged to get a final result that is more accurate and valid. Model performance can be more accurately evaluated with this method than with a single validation (holdout). More precise control over the training and validation iterations is provided by this method rather than k-fold cross-validation.

The dataset was randomly split into four parts: 60% for training, 20% for validation, 10% for test subset A, and 10% for test subset B. The proposed model was trained on the training data. After each iteration of training, the proposed models are checked against the validation set to ensure they're performing as expected. Next, we put the first ensemble layers of models through on the testing dataset A and use criteria to put a number on how well they performed. When it came time to evaluate the complete FFC/FCO/FO/FFCO-MEDNC, test subset B was put to the test. After 10 iterations, we average the results to ensure a reliable conclusion.

**3. Results**

For training purposes, we use six out of sixteen proposed classifiers as an example to train the FFC/FCO/FO-MEDNC, and eight out of sixteen proposed classifiers for the FFCO-MEDNC. Additional feature extractors and classifiers can be added to the MEDNC framework as desired.

*3.1. Results of Proposed Pre-Trained Models*

On COVID-19 dataset A, we repeated the procedure that datasets are randomly partitioned, trained, validated, and tested ten times according to MCCV and averaged the results for a more trustworthy conclusion. Table 5 displays the averaged test results of ten holdout runs for proposed models. The COVID-19 datasets are used for the training of the revised models. At the end of each training epoch, the models are checked for accuracy using a validation set. As a result of the fact that this study investigates the distinction between COVID-19 and non-COVID-19 classes, two neurons were used at the output layer [43].

3.1.1. Classification Results

From Table 5, we can see that the pre-trained models with the top-6 highest prediction accuracies are MobileNet (95.92%), ResNet101V2 (95.38%), ResNet152V2 (94.29%), MobileNetV2 (94.08%), DenseNet169 (94.02%), and DenseNet201 (93.67%). Acceptable results with over 80% accuracy were also obtained by models like InceptionResNetV2, VGG16, Xception, and VGG19. About 70% accuracy was achieved by ResNet50 and ResNet101, while 50% accuracy was provided by MobileNetV3Small. Table 6 lays out classification results for COVID-19 dataset B. It can be observed that ResNet152V2 (98.67%), ResNet101V2 (98.33%), MobileNet (98.10%), DenseNet201 (97.02%), DenseNet169 (96.74%) and MobileNetV2 (96.33%) occupied the top six highest accuracy position.

**Table 5**. The average results of applying ten iterations of modified pre-trained models to COVID-19 dataset A. Bold indicate the top 6 highest accuracies for these sixteen models.

| Model | Accuracy | Precision | Sensitivity | F1 | Specificity |
| --- | --- | --- | --- | --- | --- |
| DenseNet121 | 0.9266 | 0.9303 | 0.9402 | 0.9265 | 0.9701 |
| VGG16 | 0.8694 | 0.8854 | 0.8545 | 0.8728 | 0.8256 |
| **DenseNet201** | **0.9367** | **0.9380** | **0.9102** | **0.9384** | **0.9612** |
| ResNet50 | 0.7041 | 0.7133 | 0.7314 | 0.7269 | 0.7287 |
| **MobileNetV2** | **0.9408** | **0.9306** | **0.9502** | **0.9403** | **0.9320** |
| **ResNet152V2** | **0.9429** | **0.9440** | **0.9442** | **0.9429** | **0.9223** |
| Xception | 0.9031 | 0.9011 | 0.9265 | 0.8999 | 0.8798 |
| VGG19 | 0.9102 | 0.9130 | 0.9410 | 0.9101 | 0.8792 |
| ResNet101 | 0.6980 | 0.7216 | 0.7404 | 0.6897 | 0.6492 |
| **ResNet101V2** | **0.9538** | **0.9539** | **0.9408** | **0.9572** | **0.9613** |
| NASNet | 0.9184 | 0.9191 | 0.9388 | 0.9200 | 0.9020 |
| ResNet50V2 | 0.9069 | 0.8886 | 0.8456 | 0.8873 | 0.9931 |
| **MobileNet** | **0.9592** | **0.9583** | **0.9761** | **0.9524** | **0.9568** |
| MobileNetV3Small | 0.5000 | 0 | 0 | 0 | 0.5000 |
| InceptionResNetV2 | 0.8531 | 0.8697 | 0.9392 | 0.8514 | 0.7912 |
| **DenseNet169** | **0.9402** | **0.9481** | **0.9431** | **0.9584** | **0.9451** |

**Table 6.** The average results of applying ten iterations of modified pre-trained models to COVID-19 dataset B. Bold indicate the highest accuracy for these six models.

| Model | Accuracy | Precision | Sensitivity | F1 | Specificity |
|---|---|---|---|---|---|
| DenseNet121 | 0.9533 | 0.9541 | 0.9533 | 0.9533 | 0.9383 |
| VGG16 | 0.8820 | 0.8742 | 0.8726 | 0.8601 | 0.8849 |
| **DenseNet201** | **0.9702** | **0.9761** | **0.9730** | **0.9784** | **0.9706** |
| ResNet50 | 0.7463 | 0.7447 | 0.7549 | 0.7359 | 0.7460 |
| **MobileNetV2** | **0.9633** | **0.9606** | **0.9752** | **0.9643** | **0.9664** |
| **ResNet152V2** | **0.9867** | **0.9869** | **0.9764** | **0.9679** | **0.9811** |
| Xception | 0.9075 | 0.8605 | 0.9251 | 0.8904 | 0.8841 |
| VGG19 | 0.9241 | 0.9163 | 0.8724 | 0.8905 | 0.8694 |
| ResNet101 | 0.7639 | 0.6172 | 0.8481 | 0.7095 | 0.6917 |
| **ResNet101V2** | **0.9833** | **0.9835** | **0.9827** | **0.9889** | **0.9803** |
| NASNet | 0.9196 | 0.8703 | 0.9257 | 0.9084 | 0.8836 |
| ResNet50V2 | 0.9103 | 0.9665 | 0.8749 | 0.9251 | 0.9728 |
| **MobileNet** | **0.9810** | **0.9803** | **0.9861** | **0.9720** | **0.9807** |
| MobileNetV3Small | 0.5000 | 0 | 0 | 0 | 0.5000 |
| InceptionResNetV2 | 0.9205 | 0.8913 | 0.9324 | 0.9073 | 0.8946 |
| **DenseNet169** | **0.9674** | **0.9683** | **0.9456** | **0.9668** | **0.9685** |

### 3.1.2. Confusion Matrix Results

Figure 10 demonstrates that most of the proposed pre-trained models accurately distinguish between CT scans with and without COVID-19. With a 95.71 percent success rate, MobileNet correctly labeled 469 images. ResNet152V2 had a misclassification rate of 92.65 percent (36 of 490 instances). With a 93.47 percent success rate, DensNet201 correctly labeled 228 out of 245 images that were not part of COVID-19. Only five out of 245 COVID-19 images were not recognized by MobileNetV2 (accuracy of 92.65%).

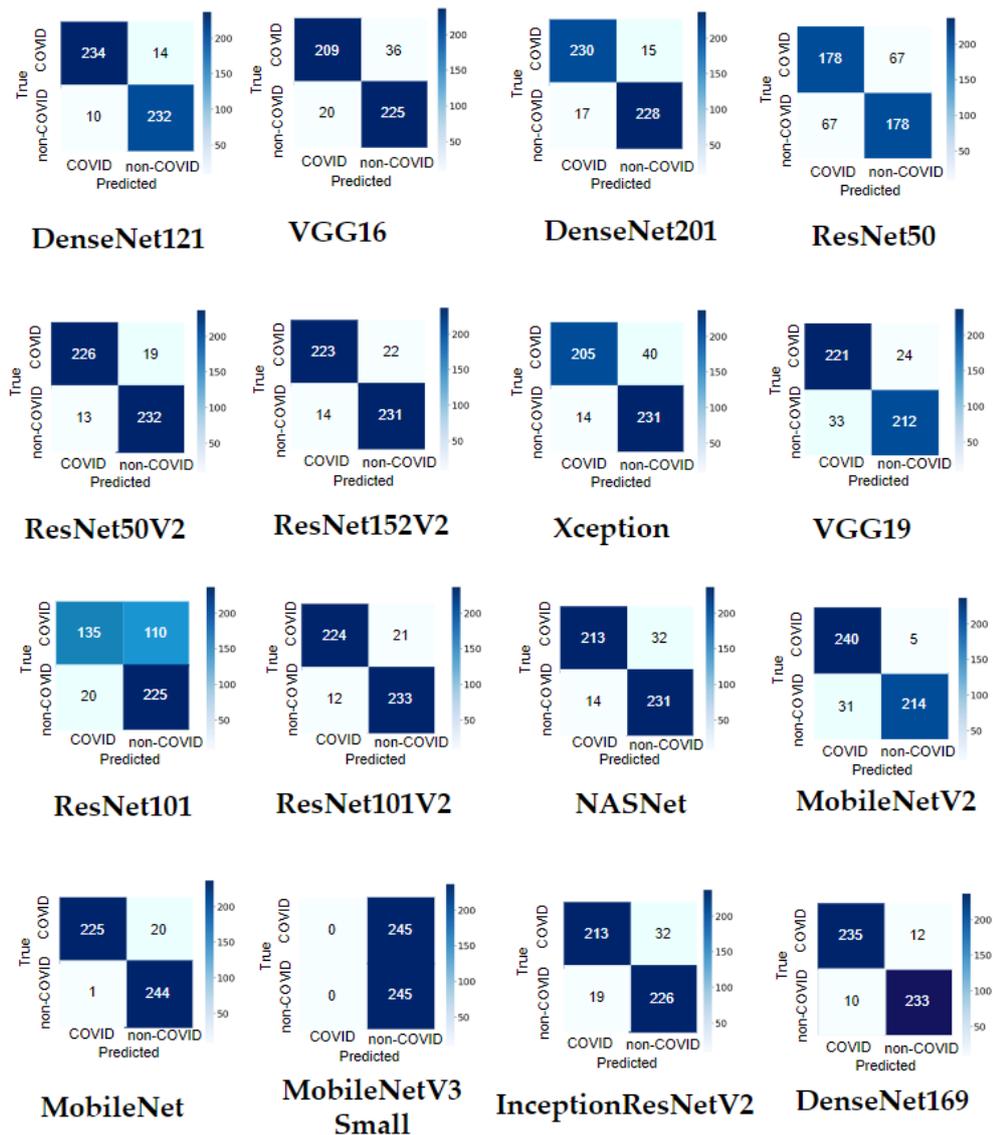

**Figure 10.** The confusion matrix for COVID-19 dataset A from a single hold-out run with 16 modified pre-trained models.

3.1.3. Learning Curve Results

Figure 11 displays the learning curves for a single run of training and validation for sixteen distinct pre-trained models. When comparing loss rates, the chart shows that MobileNet has the lowest at 10.28 percent and the highest at 95.92 percent, followed by DenseNet201 at 13.31 percent and ResNet152V2 at 20.69 percent. When compared to the training curve, all validation curves show oscillations. This is because the model's training dataset is much larger than the validation dataset, which is why this happens. A lower loss rate and higher accuracy were achieved when using validation data as opposed to training data, as depicted by the plot. This is due to the fact that we used a 0.5 dropout to get 50% of the features set to zero, during the validation phase, all neurons are utilized, leading to improved validation accuracy

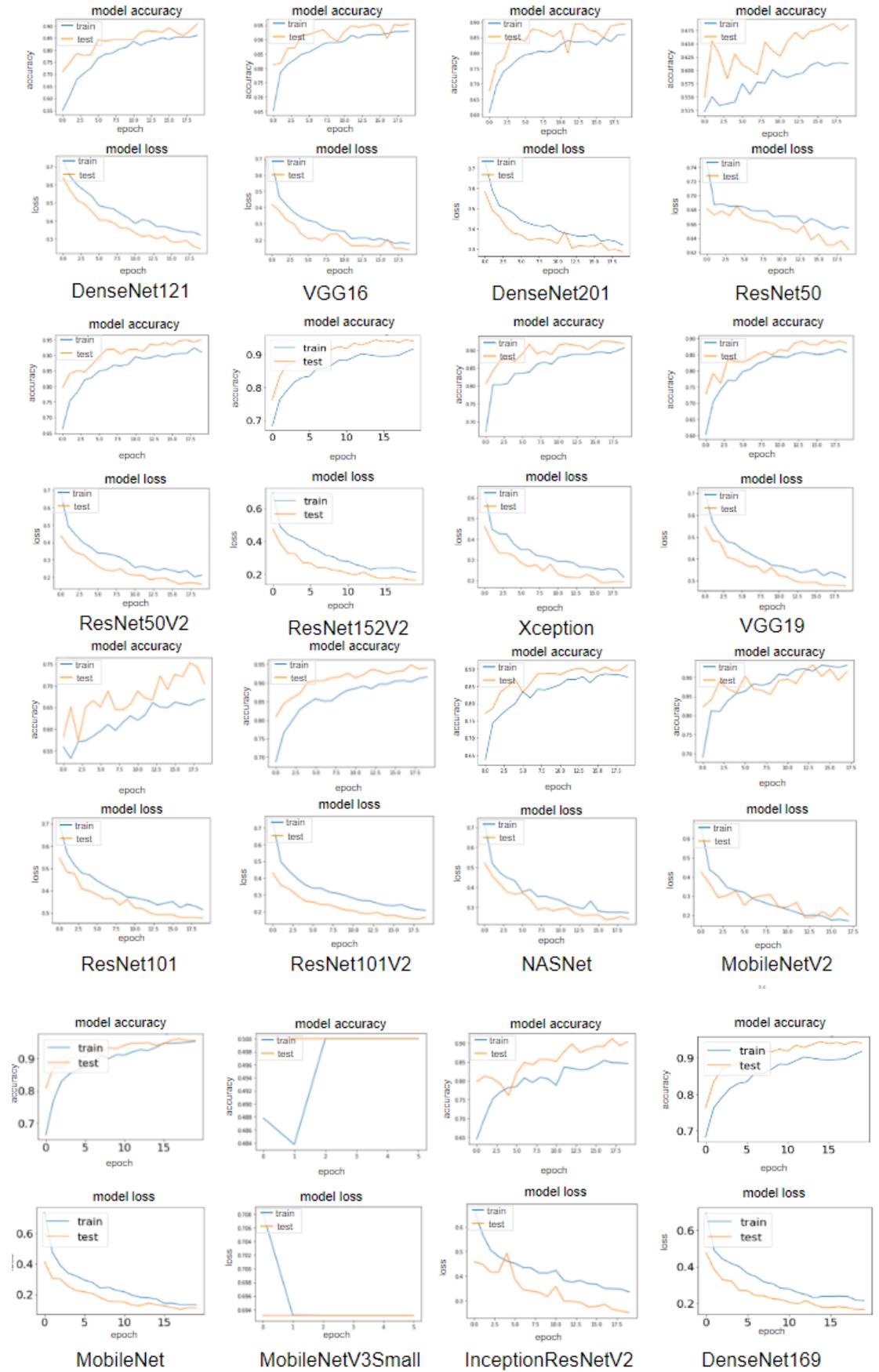

**Figure 11.** The COVID-19 dataset A learning curves for a single holdout run with sixteen differently pre-trained models.

## 3.2. Results of Proposed MEDNC

### 3.2.1. Classification Results

To identify COVID-19 CT images, four ensemble strategies have been implemented and training ten times based on MCCV. In Table 7, we can see that the all four MEDNC models have a higher accuracy rate in predicting COVID-19 dataset A than any of the individual pre-trained models. There was an increase of 2.60 percent in accuracy, 1.80 percent in precision, 0.43 percent in sensitivity, 3.69 percent in specificity, and 2.93 percent in the F1-score. The FFCO-MEDNC model has the highest accuracy (98.79%), precision (99.19%), sensitivity (98.32%), and F1-score (98.80%) of the four ensemble models.

**Table 7**. The average results of MEDNC models for running 10 times on COVID-19 dataset A. Bold suggest the highest accuracy among those models.

| Model | Accuracy | Precision | Sensitivity | F1 | Specificity |
| --- | --- | --- | --- | --- | --- |
| FFC-MEDNC | 0.9852 | 0.9763 | 0.9804 | 0.9817 | 0.9937 |
| FCO-MEDNC | 0.9758 | 0.9674 | 0.9762 | 0.9836 | 0.9829 |
| FO-MEDNC | 0.9556 | 0.9438 | 0.9571 | 0.9575 | 0.9313 |
| **FFCO-MEDNC** | **0.9879** | **0.9919** | **0.9832** | **0.9880** | **0.9918** |

Table 8 displays the results of the classification performed on the COVID-19 dataset B. The results demonstrate that the FFCO-MEDNC achieved the highest levels of accuracy (99.82%), sensitivity (99.67%), precision (99.74%), and specificity (99.89%).

**Table 8.** The average results of MEDNC models for running 10 times on COVID-19 dataset B. Bold suggest the highest accuracy among those models.

| Model | Accuracy | Precision | Sensitivity | F1 | Specificity |
| --- | --- | --- | --- | --- | --- |
| FFC-MEDNC | 0.9964 | 0.9892 | 0.9931 | 0.9947 | 0.9946 |
| FCO-MEDNC | 0.9913 | 0.9925 | 0.9924 | 0.9829 | 0.9971 |
| FO-MEDNC | 0.9906 | 0.9913 | 0.9911 | 0.9958 | 0.9872 |
| **FFCO-MEDNC** | **0.9982** | **0.9974** | **0.9967** | **0.9925** | **0.9989** |

### 3.2.2. Confusion Matrix Results

As shown in the confusion matrix in Figure 12, when compared to a single CNN model, the number of misclassifications is drastically reduced when four MEDNC models are used instead. In this single run, only six out of two hundred and forty-eight CT scans were incorrectly labeled by the FFC-MEDNC (97.58% accuracy). Classification of 237 out of 248 CT images was achieved by the FCO-MEDNC model (95.56% accuracy), while identification of 237 CT images was achieved by the FO-MEDNC model (95.56% accuracy). Only three out of two hundred and forty-eight CT scans were mislabled by the FFCO-MEDNC model (98.79% accuracy).

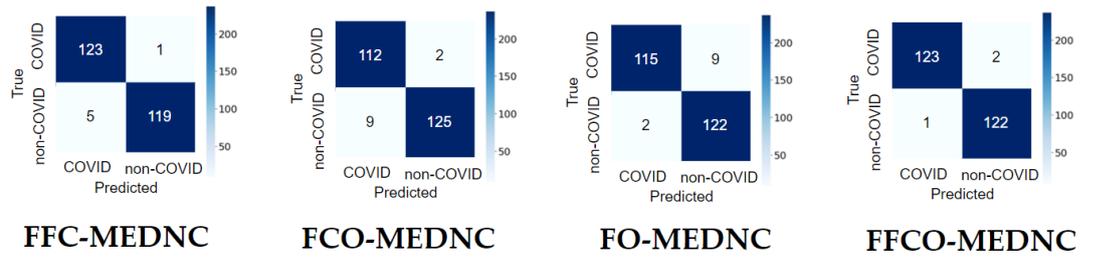

**Figure 12.** The confusion matrix for MEDNC models on COVID-19 dataset A from a single run.

3.2.3. False Discovery Rate Results

To determine the false discovery rate or FDR, divide the number of false positive test findings by the total number of positive test results. Below is the FDR formula.

$$FDR = \frac{FP}{TP + FP}$$

In Table 9, we can see that FFC-MEDNC achieved the lowest FDR (0.0081%) out of the six pre-trained models and across all ensemble models.

**Table 9**. Comparison of the FDR between the COVID-19 dataset A's pre-trained and ensemble models in a single run. The lowest FDR is highlighted in bold.

| Model | FDR |
|---|---|
| ResNet152V2 | 0.0543 |
| DenseNet201 | 0.0815 |
| MobileNet | 0.0380 |
| MobileNetV2 | 0.0462 |
| ResNet101V2 | 0.0679 |
| DenseNet169 | 0.0715 |
| **FFC-MEDNC** | **0.0081** |
| FCO-MEDNC | 0.0175 |
| FO-MEDNC | 0.0726 |
| FFCO-MEDNC | 0.0160 |

3.2.4. Learning Curve Results

Figure 13 demonstrates that compared to individually pre-trained models, MEDNC produces significantly more impressive learning curves. In comparison to FO-MEDNC, FCO-MEDNC, and FFCO-MEDNC, which have loss rates of 10.18%, 6.61%, and 1.71%, and accuracy rates of 96.82%, 97.97%, and 99.46%, respectively, FFC-MEDNC has the lowest loss rate at 1.69%. In addition, the difference between the final training and validation values is minimal in the FFCO-

MEDNC model when the training and validation loss reaches a stable stage, indicating that this model is well-fit. Differences between the training and validation values in FO-MEDNC's and loss curves are not optimistic.

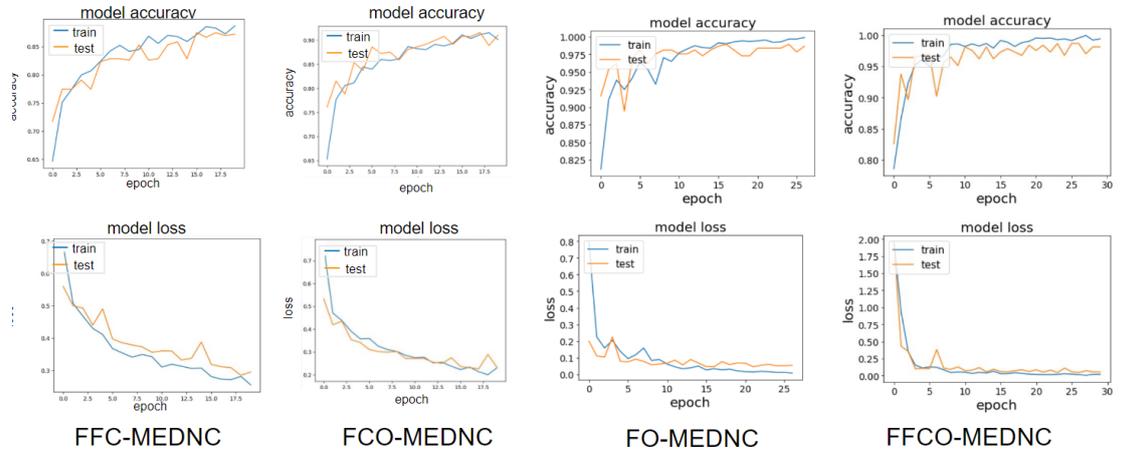

Figure 13. Learning curves for MEDNC models on COVID-19 dataset A in one hold-out run.

*3.3. Compare With the State of Art Approaches*

Twenty cutting-edge methods were chosen for analysis in this work. We found several research gaps when comparing their methods and findings to ours regarding COVID-19 recognition. Most of the studies used a dataset containing fewer than a thousand COVID-19 pictures, which is insufficient for creating effective and reliable deep learning techniques. There is a risk that the effectiveness of the proposed methods will suffer from a lack of data.

Most studies had an issue with data imbalance, where one group had more pictures than the other. As a result, model precision suffers. Secondly, some additional pre-trained models have yet to be implemented for COVID-19 classification. Moreover, in COVID-19 studies, researchers have not paid enough attention to the effects of using various ensemble methods.

In contrast, we used a CT scan dataset with a perfect balance, which included over 2,000 images. In total, sixteen different pre-trained deep learning models were examined for this project, including some that were not used in the COVID-19 recognition region. Additionally, four ensemble models for identifying COVID-19 CT images have been proposed. Table 10 is a compilation of the results from all the different models. Our research shows that the proposed model is more effective than the listed classifiers.

**Table 10.** Comparison with state-of-the-art approaches

| Author(s) | Model(s) | Optimized Model | | | |
| --- | --- | --- | --- | --- | --- |
| | | Accuracy | F1 | Recall | Precision |
| Matsuyama, E [16] | ResNet50 + wavelet coefficients | 92.2% | 91.5% | 90.4% | / |
| Loey. M[17] | ResNet50 + augmentation + CGAN | 82.91% | / | 77.66% | / |
| Do, C [18] | Modified DenseNet201 | 85% | / | 79% | 91% |
| Polsinelli, M [19] | Modified SqueezeNet | 85.03% | 86.20% | 87.55% | 85.01% |
| Panwar, H[44] | Modified VGG19 | 94.04% | | | |
| Mishra, A [45] | Customized DenseNet201, VGG16, ResNet50, InceptionV3 and DenseNet121 | 88.3% | 86.7% | | 90.15% |
| Ko. H [20] | Customized Xception, ResNet50, VGG16, and Inception-v3 | 96.97% | | | |
| Maghdid.H [21] | Customized CNN, Alexnet | 94.1% | | 100% | |
| Arora.V [46] | Customized XceptionNet, MobileNet, ResNet50, DenseNet121, InceptionV3, VGG16 | 94.12% | 96.11% | 96.11% | 96.11% |
| Alshazly. H [47] | CovidResNet and CovidDenseNet | 93.87% | 95.70 | 92.49 | 99.13% |
| Yu, Z [48] | Customized DenseNet201, InceptionV3, ResNet101, ResNet50 | 95.34% | | | |
| Jaiswal, A [49] | Modified DenseNet201 | 96.25% | 96.29% | 96.29% | 96.29% |
| Sanagavarapu.S [50] | Ensembled ResNets | 87% | 84% | 81% | 91% |
| Song, J [51] | A large-scale bi-directional generative adversarial network | | | 92% | |
| Sarker, L [52] | Modified Densenet121 | 96.49% | 96% | 96% | 96% |
| Shan, F [53] | VB-Net | 91.6% | | | |
| Wang, S [54] | Modified DenseNet | 85% | 90% | 79% | |
| Gozes, O [55] | Modified ResNet50 | | | 94% | |
| Wang, S [56] | Modified Inception | 79.3% | 63% | 83% | |
| Li, L [30] | Modified RestNet50 | | | 90% | |
| Proposed | FFCO-MEDNC | 98.79% | 98.80% | 98.32% | 99.19% |

### 3.4 Generality of Proposed Method

To demonstrate the generalizability of our model, we apply our proposed model to two additional datasets, one brain tumor dataset, and one blood cell dataset.

#### 3.4.1 Brain Tumor Dataset

Scanning the brain with radiation is one approach to finding malignant growths[57]. The process of taking an X-ray, which is used to obtain images of the inside of the body, is quick and causes no discomfort to the patient[58]. The Brain Tumor Dataset[59] (JPG format) is an X-ray dataset used for this brain tumor recognition task. There were a confirmed 2341 cases of brain tumors among these images, while the remaining 2087 were deemed to be healthy. To ensure that the data is precisely balanced, we chose 2087 images to represent each category.

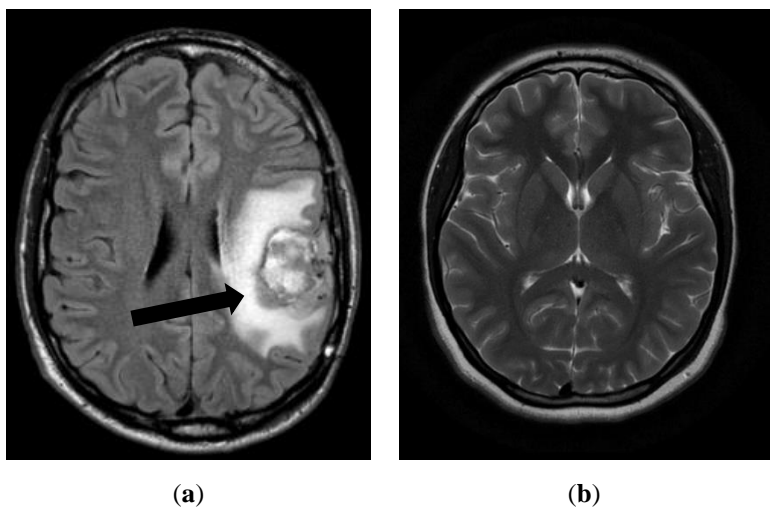

(**a**)        (**b**)

**Figure 14.** Patients with and without brain tumors in their X-ray scans. (a) An X-ray of a subject with a brain tumor. (b) An X-ray of a healthy subject.

**Table 11.** Information about the Brain Tumor dataset that was chosen

| Class | Quantities of Samples | Format |
|---|---|---|
| Brain tumor | 2087 | JPG |
| Healthy | 2087 | JPG |

3.4.2 Blood Cell Dataset

Since white blood cells (WBC) develop immunity to fight against pathogens and foreign chemicals, it is vital to distinguish the various WBC subsets. Accordingly, we use our proposed models to complete the WBC classification. Specifically, 12,500 JPEG pictures of a blood cell dataset[60] taken using a microscope from one of three different cell types are being used for this purpose. For each of the three types of cells studied, we randomly picked 2497 images. The three kinds of cells are eosinophils, lymphocytes, and monocytes.

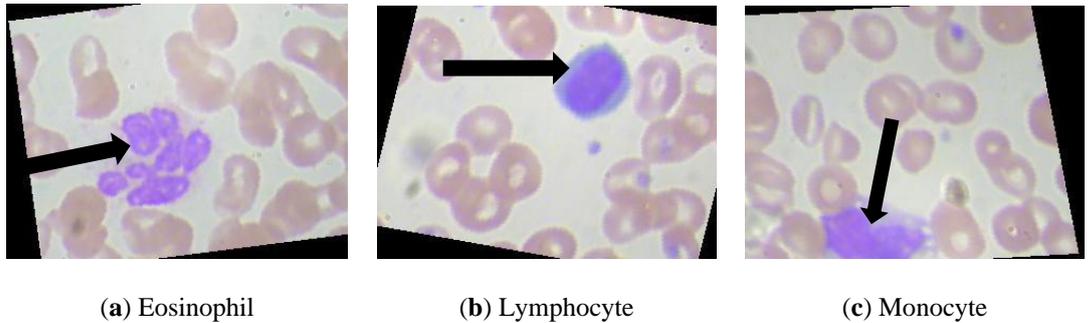

(**a**) Eosinophil        (**b**) Lymphocyte        (**c**) Monocyte

**Figure 15.** Blood cell dataset. (a) Eosinophil cells. (b) Lymphocyte cells. (c) Monocyte cells.

**Table 12.** Details about the SARS-CoV-2 Chest CT scan dataset that was chosen

| Class | Numbers of Samples | Format |
|---|---|---|
| Eosinophil | 2497 | JPEG |
| Lymphocyte | 2497 | JPEG |
| Monocyte | 2497 | JPEG |

We tested our MEDNC (FFC-MEDNC, FCO-MEDNC, FO-MEDNC, and FFCO-MEDNC) on these two datasets, and the results below showed the adaptability of our model towards other medical images other than COVID-19.

3.4.3 Brain Tumor Results

The six pre-trained models with the highest prediction accuracy, according to Table 13, are MobileNet (98.87%), ResNet101V2 (98.08%), ResNet152V2 (97.92%), MobileNetV2 (96.42%), DenseNet169 (96.81%), and DenseNet201 (96.17%). The accuracy of the four MEDNC models in predicting brain tumor dataset is more than that of any of the individual pre-trained models, as shown in Table 14. Accuracy increased by 0.41 percent, precision by 0.51 percent, sensitivity by 0.91 percent, specificity by 1.23 percent, and F1-score by 0.56 percent. The FFCO-MEDNC model has the greatest accuracy (99.39%), sensitivity (99.72%), precision (99.41%), and F1 score

(99.53%) among the three ensemble models.

**Table 13**. The average results of 10 runs of the six modified pre-trained models with the highest accuracy utilizing brain tumor data.

| Model | Accuracy | Precision | Sensitivity | F1 | Specificity |
|---|---|---|---|---|---|
| DenseNet169 | 0.9681 | 0.9874 | 0.9615 | 0.9679 | 0.9910 |
| DenseNet201 | 0.9617 | 0.9705 | 0.9629 | 0.9598 | 0.9865 |
| ResNet152V2 | 0.9792 | 0.9710 | 0.9728 | 0.9804 | 0.9839 |
| ResNet101V2 | 0.9808 | 0.9811 | 0.9894 | 0.9740 | 0.9934 |
| MobileNet | 0.9887 | 0.9862 | 0.9847 | 0.9785 | 0.9935 |
| MobileNetV2 | 0.9642 | 0.9553 | 0.9613 | 0.9748 | 0.9711 |

**Table 14**. The average results of 10 runs of MEDNC models using brain tumor data are as follows.

| Model | Accuracy | Precision | Sensitivity | F1 | Specificity |
|---|---|---|---|---|---|
| FFC-MEDNC | 0.9928 | 0.9864 | 0.9985 | 0.9927 | 0.9991 |
| FCO-MEDNC | 0.9907 | 0.9925 | 0.9914 | 0.9885 | 0.9953 |
| FO-MEDNC | 0.9856 | 0.9799 | 0.9847 | 0.9813 | 0.9756 |
| **FFCO-MEDNC** | **0.9939** | **0.9941** | **0.9972** | **0.9953** | **0.9981** |

3.4.4 Blood Cell Results

According to Table 15, the six pre-trained models with the highest prediction accuracy are MobileNet (96.77%), MobileNetV2 (93.05%), DenseNet169 (92.38%), DenseNet201 (90.76%), ResNet101V2 (89.37%), and ResNet152V2 (86.75%). As shown in Table 16, the accuracy of the four MEDNC models in predicting the blood cell dataset is higher than that of any of the individual pre-trained models. Accuracy improved by 2.26 percent, precision improved by 3.60 percent, sensitivity improved by 6.83 percent, specificity improved by 2.21 percent, and F1-score improved by 3.00 percent. Among the three ensemble models, the FFCO-MEDNC model has the highest accuracy (99.28%), sensitivity (99.84%), and F1 score (99.52%). The FCO-MEDNC has the highest precision (99.63%) and specificity (99.67%)

**Table 15**. The following are the average results of 10 runs of the six modified pre-trained models with the highest accuracy utilizing the blood cell dataset.

| Model | Accuracy | Precision | Sensitivity | F1 | Specificity |
|---|---|---|---|---|---|
| DenseNet169 | 0.9238 | 0.9220 | 0.9183 | 0.9188 | 0.9383 |
| DenseNet201 | 0.9076 | 0.9026 | 0.8967 | 0.8974 | 0.9168 |
| ResNet152V2 | 0.8675 | 0.8556 | 0.8550 | 0.8535 | 0.9180 |
| ResNet101V2 | 0.8937 | 0.8140 | 0.8663 | 0.8088 | 0.9137 |
| MobileNetV2 | 0.9305 | 0.9236 | 0.9175 | 0.9190 | 0.9762 |
| MobileNet | 0.9677 | 0.9603 | 0.9784 | 0.9667 | 0.9242 |

**Table 16.** The following are the average results of 10 runs of the six modified pre-trained models with the highest accuracy utilizing the blood cell dataset.

| Model | Accuracy | Precision | Sensitivity | F1 | Specificity |
|---|---|---|---|---|---|
| FFC-MEDNC | 0.9903 | 0.9963 | 0.9858 | 0.9946 | 0.9983 |
| FCO-MEDNC | 0.9896 | 0.9842 | 0.9869 | 0.9713 | 0.9927 |
| FO-MEDNC | 0.9824 | 0.9915 | 0.9794 | 0.9925 | 0.9863 |
| **FFCO-MEDNC** | **0.9928** | **0.9926** | **0.9984** | **0.9952** | **0.9967** |

*3.5. Training Duration and Model Size Outcomes*

From Table 17, we can see that MobileNet required the least length of time to finish one epoch of training. However, NASNet was the slowest, taking 32 seconds to complete one epoch of training. Table 17 also displays the sizes of the models. The proposed FFCO-MEDNC has the largest model size at 392.5 MB, while the smallest, MobileNetV3Small, is only 13.26 MB.

**Table 17.** A comprehensive evaluation of all models' training times and weights for the COVID-19 dataset A.

| Model | Accuracy | Training Time (Second/Epoch) | Parameters (MB) |
|---|---|---|---|
| ResNet50 | 0.7659 | 27 | 103.99 |
| ResNet101V2 | 0.9506 | 29 | 177.08 |
| ResNet152V2 | 0.9456 | 28 | 237.5 |
| ResNet50V2 | 0.9286 | 26 | 103.9 |
| ResNet101 | 0.7302 | 28 | 177.19 |
| VGG16 | 0.8728 | 26 | 63.08 |
| VGG19 | 0.8746 | 26 | 79.87 |
| DenseNet121 | 0.9304 | 29 | 83.28 |
| Densenet169 | 0.9483 | 29 | 86.31 |
| DenseNet201 | 0.9349 | 30 | 84.37 |
| InceptionResNetV2 | 0.8971 | 30 | 213.77 |
| Xception | 0.8978 | 27 | 93.48 |
| NASNet | 0.9108 | 32 | 25.26 |
| MobileNet | 0.9536 | 25 | 19.33 |
| MobileNetV2 | 0.9465 | 27 | 17.52 |
| MobileNetV3Small | 0.5000 | 28 | 13.26 |
| FFC-MEDNC (Ours) | 0.9852 | 31 | 356.4 |
| FCO-MEDNC (Ours) | 0.9758 | 31 | 346.39 |
| FO-MEDNC (Ours) | 0.9556 | 31 | 332.7 |
| FFCO-MEDNC(Ours) | 0.9879 | 31 | 392.5 |

**4. Conclusion**

The use of deep learning techniques for diagnosing COVID-19 has recently become a topic of intense interest. Exciting progress has been made, and new insights continue to emerge, thanks to the use of multiple neural networks in this area of study. Among the most important forms of data used to detect COVID-19 symptoms are CT scan images. With the purpose of identifying COVID-19, a plethora of deep learning models have been built and effectively applied. Using chest CT scans as training data, this paper adapts and creates sixteen deep-learning models for recognizing COVID-19. As a follow-up, MEDNC (FFC-MEDNC, FCO-MEDNC, FO-MEDNC, and FFCO-MEDNC) were proposed to improve the performance of screening COVID-19 using the aforementioned two COVID-19 CT scan datasets. The MEDNC models were found to be superior to the customized pre-trained models on the COVID-19 image recognition task.

The findings indicate that FFCO-MEDNC obtained an accuracy of 98.79%, a sensitivity of 98.32%, a precision of 99.19%, and a specificity of 99.18% for COVID-19 dataset A. COVID-19 dataset B improved to 99.82 percent accuracy rate, 99.67 percent sensitivity, 99.74 percent precision, and 99.89 percent specificity.

## 5. Feature Works

Although the proposed COVID-19 recognition system performed exceptionally well, there were still some caveats to this study. Before anything else, the classification outcome may change depending on which 2D image is chosen when deriving a 3D image from a CT scan. Second, additional preprocessing techniques, such as image enhancement, have not been incorporated into this investigation. Image enhancement software could be used to check if there is room for improvement in future work. These results suggest that a fully automated and rapidly performed diagnosis of COVID-19 using deep learning is possible with the help of the proposed MEDNC models. This discovery will help doctors save both time and money in their efforts to detect COVID-19 infections.

**Author Contributions:**

Lin Yang: Conceptualization, Software, Validation, Investigation, Data Curation, Writing - Original Draft, Visualization

Shuihua Wang: Methodology, Formal analysis, Resources, Data Curation, Writing - Review & Editing, Supervision, Funding acquisition

Yudong Zhang: Conceptualization, Methodology, Formal analysis, Investigation, Writing - Review & Editing, Supervision, Project administration,

**Funding:** This paper is partially supported by Medical Research Council, UK (MC_PC_17171); Royal Society, UK (RP202G0230); BHF Accelerator Award, UK (AA/18/3/34220); Hope Foundation for Cancer Research, UK (RM60G0680); GCRF, UK (P202PF11); Sino-UK Industrial Fund, UK (RP202G0289); LIAS, UK (P202ED10, P202RE969); Data Science Enhancement Fund, UK (P202RE237); Fight for Sight, UK (24NN201); Sino-UK Education Fund, UK (OP202006); BBSRC, UK (RM32G0178B8).

**Data Availability Statement:** The two datasets can be accessed from https://www.kaggle.com/plameneduardo/sarscov2-ctscan-dataset and https://www.kaggle.com/hgunraj/covidxct

**Conflict of Interest:** The authors declare no conflict of interest.